\documentclass[letterpaper,11pt]{article}
\pdfoutput=1 

\usepackage{jheppub} 

\usepackage{mathtools,enumerate}
\usepackage{multirow,array,romannum}

\title{\boldmath Thermofield Duality for Higher Spin Rindler Gravity}

\author{Antal Jevicki,}
\author{Kenta Suzuki}

\affiliation{Department of Physics, Brown University,\\Providence, RI 02912, USA}

\emailAdd{Antal\_Jevicki@brown.edu}
\emailAdd{Kenta\_Suzuki@brown.edu}

\preprint{{\tt BROWN-HET-1676}}

\abstract{
We study the Thermo-field realization of the duality between the Rindler-AdS higher spin theory and $O(N)$ vector theory.
The CFT represents a decoupled pair of free $O(N)$ vector field theories.
It is shown how this decoupled domain CFT is capable of generating the connected Rindler-AdS background with the full set of Higher Spin fields.
}

\begin{document} 

\maketitle
\flushbottom

\section{Introduction}
\label{sec:introduction}
An important investigation in AdS/CFT Duality involves Black Hole (BH) spacetimes, or spacetimes with horizons
\cite{Mathur:2012np, Braunstein:2009, Almheiri:2012rt, Marolf:2013dba, VanRaamsdonk:2010pw}.
The CFT dual in these cases is to be represented by Large $N$ field theories at finite temperature.
The simplest example of such duality is the Rindler version of AdS \cite{Czech:2012, Parikh:2012} (also found as the massless topological Black Hole \cite{Emparan:1999, Aros:2002}).
In this case, one has two complementary Rindler wedges of the AdS space and the dual description is to be constructed in terms of a pair of decoupled CFT's \cite{Maldacena:2001kr}.
At finite temperature, this system represents a system known as the ``thermo-field double''
\cite{Israel:1976ur, Schwinger:1960qe, Keldysh:1964ud, Semenoff:1982ev, Niemi:1983nf, Ojima:1981ma}.
There are general discussions in the literature concerning the question of how two decoupled CFT's are able to reconstruct a connected space-time of a black hole or even a Rindler-AdS
\cite{Maldacena:2013xja, Avery:2013bea, Mathur:2014dia}.
An application of this to gravity on noncommutative spaces was recently developed in \cite{Nair:2015}.
Arguments in favor of such a construction in the case of Rindler Gravity were given in \cite{Czech:2012, Parikh:2012}, (see also \cite{Bousso:2012mh}).
In this work we will be considering generalized Higher Spin \cite{Fradkin:1986ka, Vasiliev:1990en, Vasiliev:1995dn, Bekaert:2005vh}
dualities \cite{Klebanov:2002ja, Sezgin:2002rt} in ${\rm AdS}_4$
which have been intensively studied during the last few years
\cite{Giombi:2009wh, Koch:2010cy, Maldacena:2011jn, Kraus:2011ds, Gaberdiel:2012yb, Ammon:2012wc, Gaberdiel:2013jca, Das:2003vw, Koch:2014aqa, Mintun:2014gua,
Leigh:2014qca, Jin:2015aba, Koch:2014mxa, Jevicki:2015sla}.
They involve $O(N)$ vector models at their critical points of which the free theory is a particularly simple and tractable example.
For pure AdS, it allows for a complete reconstruction of the bulk theory with higher spin fields and to all orders in $1/N$.
A direct map between CFT observables and bulk AdS fields was shown to be given in terms of a bi-local composite field of the $O(N)$ CFT \cite{Koch:2010cy, Koch:2014aqa}.
A study of this bi-local construction at finite temperature in the Thermo-field formulation was given recently in \cite{Jevicki:2015sla}
with results showing the characteristics of bulk black hole spacetimes.
In this paper, we sharpen this correspondence by giving a construction of Rindler-AdS Higher Spin Gravity.
In this case one has a pair of CFT's on hyperbolic spaces that are to be quantized in the Thermo-field scheme.
In addition to presenting a more complete example of ``domain duality'' \cite{Bousso:2012mh}
and the ability of reconstructing a connected spacetime, we clarify the stability of this system.
The instability of the quotient hyperbolic space case was studied in \cite{Belin:2014}.

The content of this paper is as follows: In Section~\ref{sec:massless limit of ads bh and rindler-ads},
we review the appearance of asymptotic Rindler-AdS spacetime as the massless limit of particular black holes (Topological Black Holes).
We also give some relevant properties of this spacetime such as the presence of ``evanescent'' modes \cite{Rey:2014dpa} whose CFT reconstruction will be of particular interest.
In Section~\ref{sec:o(n) vector model on hyperbolic space}, we discuss the quantization of the $O(N)$ vector model on hyperbolic space and its hamiltonian Thermo-field version.
In Section~\ref{sec:bi-local mapping}, we give details of our bi-local map to Rindler-AdS bulk.
Demonstration of ``evanescent'' modes given by this construction is one application.
Sect 5 is reserved for Conclusions regarding the more general questions involving reconstruction of spacetime behind the horizon of BH
\cite{Papadodimas:2012aq, Papadodimas:2013jku, Kraus:2002iv}.

\section{Massless Limit of AdS BH and Rindler-AdS}
\label{sec:massless limit of ads bh and rindler-ads}
In this section, we review the static black hole solutions in asymptotically AdS spacetime and their massless limit, mainly following the discussion of \cite{Emparan:1999}.

Asymptotically AdS space admits three kinds of the static back hole solutions.
These three solutions are characterized by the curvature of the horizons (transverse spacial coordinates).
For AdS$_{d+1}$, the metric is explicitly given by
	\begin{align}
		ds^2 \, = \, - \, f_k(r) \, dt^2 \, + \, \frac{dr^2}{f_k(r)} \, + \, \frac{r^2}{L^2} \, d\Sigma_{k, d-1}^2 \, ,
	\label{ads-bh}
	\end{align}
where
	\begin{align}
		f_k(r) \, = \, \frac{r^2}{L^2} \, - \, \frac{\mu}{r^{d-2}} \, + \, k \, .
	\end{align}
Here $L$ denotes the AdS radius and $\mu$ is a deformation parameter, which is proportional to the black hole mass.
$d\Sigma_{k, d-1}^2$ represents the metric of the $(d-1)$-dimensional horizon given by 
the unit metric of ($S^{d-1}$, $\mathbb{R}^{d-1}$, $H_{d-1}$) for $k=(+1, 0, -1)$ case respectively.

The positive horizon curvature solution ($k=+1$) is called ``AdS-Schwarzschild'' black hole,
the zero curvature solution ($k=0$) is ``AdS-Planar'' black hole,
and the negative curvature ($k=-1$) is ``AdS-Hyperbolic'' black hole.
The horizon location $r_+$ is determined by the larger solution of $f_k(r_+)=0$.
In the massless limit $\mu \to 0$, each black hole solution is reduced to the global, Poincare, and Rindler coordinates of AdS spacetime, respectively.
The temperature of these black holes is determined as
	\begin{align}
		\beta \, = \, \frac{4 \pi L^2 r_+}{r_+^2 d + k(d-2) L^2} \, .
	\end{align}

The boundary metric is given by taking $r \to \infty$.
Therefore, one expects that each AdS black hole solution is dual to CFT's on
$\mathbb{R}\times S^{d-1}$, $\mathbb{R}^{1,d-1}$, and $\mathbb{R}\times H^{d-1}$, respectively \cite{Emparan:1999}.
Hereafter, we set $L=1$.

In this paper, we are mainly interested in the four-dimensional Rindler-AdS coordinates (this is sometimes called massless topological black hole),
which is given as the $k=-1$ and $\mu=0$ case of (\ref{ads-bh}).
Note that for this case, the Rindler horizon location is given by $r_+=1$ and $\beta=2\pi$, which agrees with the Unruh temperature.
For our purpose, it's more convenient to use a Poincare-like radial coordinate $\rho$
\footnote{
Usually, $\rho$ is used for the global-like radial coordinate; however, in this paper we use it for the Poincare-like coordinate.
This unusual notation is because when we construct the bi-local map in Section~\ref{sec:bi-local mapping}, we will use $z$ for the Poincare radial coordinate.
}
defined by $\rho=r^{-1}$, rather than the global-like coordinate $r$.
Then, our Rindler-AdS metric is defined by
	\begin{align}
		ds^2 \, = \, \frac{1}{\rho^2} \left[ \, - (1-\rho^2) d\tau^2 \, + \, \frac{d\rho^2}{1-\rho^2} \, + \, \frac{dx^2+d\sigma^2}{\sigma^2} \, \right] \, .
	\label{Rindler-AdS}
	\end{align}
The range of each coordinate is 
	\begin{align}
		- \infty < \tau , \, x< \infty \, , \qquad 0 < \rho < 1 \, \qquad \sigma > 0 \, .
	\label{region}
	\end{align}
Note that now the boundary is located at $\rho=0$, while the Rindler horizon is at $\rho=1$.

One of the important characters of the black hole background is the presence of the so called ``evanescent modes'' \cite{Rey:2014dpa, Bousso:2012mh}.
In \cite{Rey:2014dpa}, the authors described that the evanescent modes exist in the AdS-Schwarzschild black hole background
and in its infinite volume limit, AdS-Planar black hole background.
The existence of the evanescent modes in Rindler-AdS background was suggested in the discussion of smearing function \cite{Hamilton:2006, Bousso:2012mh},
and in Appendix~\ref{sec:evanescent modes in rindler-ads}, we will demonstrate the evanescent modes in this background from the point of view of the effective potential.

Israel \cite{Israel:1976ur} described that observers in a spacetime limited by an event horizon detect a thermal spectrum of particles. 
This discussion unifies the Hawking radiation \cite{Hawking} and the Unruh effect \cite{Unruh} in the language of ``thermofield dynamics'' \cite{Takahashi:1996zn}.
Namely, one can expand the field only in the right wedge of the Kruskal coordinates (right Rindler wedge) for a black hole background (Rindler spacetime)
using the complete set of the eigenfunctions of the Klein-Gordon equation with the right oscillator $a(\vec{k})$ and its hermitian conjugate.
Here, $\vec{k}$ denotes the necessary amount of quantum numbers which we need to distinguish the eigenfunctions; we do not need a specific definition here. 
One can expand the field in the same way in the left wedge using the left oscillator $\tilde{a}(\vec{k})$ and its conjugate.
Then, from these two kinds of oscillators, one can construct a natural ``thermal'' vacuum.
Now respect to the thermal vacuum, the expectation value of any operator produces the correct thermal ensemble average.

Based on this Israel's description of the thermal nature of black holes,
Maldacena \cite{Maldacena:2001kr} proposed a two entangled CFT's description of asymptotically AdS eternal black holes in the context of the AdS/CFT correspondence. 
Namely, the Hartle-Hawking state in an eternal big AdS-Schwarzschild black hole can be viewed from the entangled two CFT's as
	\begin{align}
		| \Psi \rangle \, = \, \frac{1}{\sqrt{Z(\beta)}} \sum_n e^{-\beta E_n / 2} | E_n \rangle_1 \otimes | E_n \rangle_2 \, ,
	\label{entangled state}
	\end{align}
where $| E_n \rangle_1$ is the energy eigenstate of the ``right boundary'' CFT on $\mathbb{R}\times S^{d-1}$ and $| E_n \rangle_2$ is for the left.
The summation is taken over all energy eigenstates.
Then, the Maldacena's proposal is that a state in the eternal AdS-Schwarzschild background is dual to
the particularly entangled state (\ref{entangled state}) of the two copies of the boundary CFT's.
From the pair CFT's point of view $\beta$ is a parameter to characterize ``how much'' the pair CFT's are entangled each other.
On the other hand, in AdS point of view, this is the inverse temperature of the black hole. 

Following Maldacena's conjecture, the authors of \cite{Czech:2012} proposed a dual description of states in Rindler-AdS
from the particularly entangled pair of CFT's living on $\mathbb{R}\times H^{d-1}$.
For this conjecture, one uses exactly the same expression of the entangled state as (\ref{entangled state})
except for that $\beta$ is now taken to be the Unruh temperature $\beta=2\pi$.
As in the regular Rindler space (Rindler description of the Minkowski space) case, where we can obtain a significant amount of insight about black hole physics,
we would expect that this Rindler-AdS/CFT correspondence is a very helpful example to study the nature of black hole AdS/CFT correspondence.
This also provides a specific example to investigate about subregion dualities \cite{Bousso:2012mh}.

\section{$O(N)$ Vector Model on Hyperbolic Space}
\label{sec:o(n) vector model on hyperbolic space}
As a concrete example of the AdS/CFT correspondence, in this paper we employ the minimal bosonic higher spin and free $O(N)$ vector model duality \cite{Klebanov:2002ja}.
In this section, we define the dual CFT of Rindler-AdS higher spin theory.
It is also known that the dual Vasiliev's Higher Spin gravity possess a class of Black Hole type solutions \cite{Didenko:2009td}.
Here we clarify the relation between the hyperbolic and Rindler $O(N)$ vector models and demonstrate the stability in these models.
We will present details of canonical quantization in the Rindler spacetime in Appendix~\ref{sec:canonical quantization of rindler o(n) vector model},
which will be extensively used to discuss this Rindler higher spin/vector model duality
because this is more convenient than the vector model on hyperbolic space.

Let us first define the corresponding CFT to the Rindler-AdS (\ref{Rindler-AdS}).
By taking boundary limit ($\rho \to 0$) of the Rindler-AdS metric, one obtains
	\begin{align}
		ds^2 \, = \, - \, d\tau^2 \, + \, \frac{dx^2+d\sigma^2}{\sigma^2} \, .
	\label{hyperbolic-cft}
	\end{align}
Therefore, in \cite{Czech:2012, Parikh:2012} it was suggested that Rindler-AdS is dual to the CFT on $\mathbb{R} \times H^2$.
For our case, the vector model is defined by
	\begin{align}
		S \, = \, - \, \frac{1}{2} \int d\tau dx d\sigma \sqrt{-g} \, \Big[ \, g^{\mu\nu} \partial_{\mu} \phi_i^H \partial_{\nu} \phi_i^H \, + \, \frac{R}{8} \, \phi_i^H \phi_i^H \, \Big] \, ,
	\end{align}
where the metric is given in (\ref{hyperbolic-cft}) and the Ricci scalar is $R=-2$.
Explicitly, the action is given by
	\begin{align}
		S \, = \, \frac{1}{2} \int d\tau dx d\sigma \left[ \, \frac{1}{\sigma^2} \, (\partial_{\tau} \phi_i^H)^2 \, - \, (\partial_x \phi_i^H)^2
		\, - \, (\partial_{\sigma} \phi_i^H)^2 \, + \, \frac{1}{4\sigma^2} \, \phi_i^H \phi_i^H \, \right] \, .
	\label{action-hyper}
	\end{align}

The claim of \cite{Czech:2012, Parikh:2012} is absolutely correct; however, it's more convenient to move to the Rindler coordinates rather than hyperbolic space.
Namely, the theory is conformally invariant and the metric (\ref{hyperbolic-cft}) is by the Weyl transformation related to the regular Rindler metric
	\begin{align}
		\qquad ds^2 \, = \, - \, \sigma^2 \, d\tau^2 \, + \, dx^2 \, + \, d\sigma^2 \, , \qquad (R \, = \, 0) \, .
	\label{rindler-cft}
	\end{align}
On this background, the $O(N)$ vector model is given by
	\begin{align}
		S \, =& \ - \, \frac{1}{2} \int d\tau dx d\sigma \sqrt{-g} \, g^{\mu\nu} \partial_{\mu} \phi_i^R \partial_{\nu} \phi_i^R \nonumber\\
		=& \ \frac{1}{2} \int d\tau dx d\sigma \left[ \, \frac{1}{\sigma} \,
		(\partial_{\tau} \phi_i^R)^2 \, - \, \sigma (\partial_x \phi_i^R)^2 \, - \, \sigma (\partial_{\sigma} \phi_i^R)^2 \, \right] \, .
	\label{action-rindler}
	\end{align}
One can see these two actions (\ref{action-hyper}), (\ref{action-rindler}) are related by a field redefinition
	\begin{gather}
		\phi_i^H \, = \, \sqrt{\sigma} \, \phi_i^R \, ,
	\end{gather}
up to the total derivative term.

Note that the hyperbolic action (\ref{action-hyper}) looks effectively tachyonic and unstable.
However, one can see that this theory is not unstable just looking at the Rindler action (\ref{action-rindler}).

We give a detail discussion of canonical quantization for the $O(N)$ vector model in Rindler spacetime in Appendix~\ref{sec:canonical quantization of rindler o(n) vector model}.
Here, we give the most important results.
Starting from $3d$-Minkowski space $ds^2=-dt^2+dx^2+dy^2$, the right Rindler wedge is constructed by the following coordinate transformations
	\begin{align}
	\begin{cases} 
		\ t \, = \, \sigma \, \sinh \tau \, , \\
		\, y \, = \, \sigma \, \cosh \tau \, ,
	\end{cases}
	\label{coor-trans1}
	\end{align}
which induces the regular Rindler metric (\ref{rindler-cft}).
The regions of the new coordinates are $-\infty < \tau < \infty,\ \sigma>0$.

The solution for the equation of motion of the Rindler field with an appropriate normalization is given by
	\begin{align}
		\phi_i^R(\tau, x, \sigma) \, = \, \int_0^{\infty} d\omega \int_{-\infty}^{\infty} dk \Big[ \ b_i^R(\omega, k) \, g^R_{\omega, k}(\tau, x, \sigma)
		\, + \, b_i^{R\dagger}(\omega, k) \, g^{R*}_{\omega, k}(\tau, x, \sigma) \ \Big] \, ,
	\label{solution-r}
	\end{align}
where
	\begin{align}
		g^R_{\omega, k}(\tau, x, \sigma) \, = \, \frac{1}{2\pi \sqrt{2\omega}} \, \frac{2K_{i \omega}(|k| \sigma)}{|\Gamma(i\omega)|} \, e^{i k x - i \omega \tau} \, .
	\end{align}

From the action (\ref{action-rindler}), the Hamiltonian is defined by
	\begin{align}
		H \, = \, \int_{\Sigma} d\sigma dx \, \frac{\sigma}{2} \Big[ \, \Pi_i^2 \, + \, (\partial_{\sigma} \phi_i)^2 \, + \, (\partial_x \phi_i)^2 \, \Big] \, ,
	\end{align}
where $\Sigma$ is a constant $\tau$ hypersurface, and $\Pi_i$ is the conjugate momentum defined by $\Pi_i=\sigma^{-1} \partial_{\tau} \phi_i$.
Using the solution (\ref{solution-r}), one can rewrite the Hamiltonian in terms of the right Rindler oscillators as
	\begin{align}
		H \, =& \ \int d\omega dk \, \frac{\omega}{2} \bigg[ b_i^R(\omega, k) \, b_i^{R\dagger}(\omega, k) \, + \, {\rm h.c.} \ \bigg] \, .
	\label{hamiltonian}
	\end{align}
We give the detail of this calculation in Appendix~\ref{sec:hamiltonian of rindler vector model}.

\section{Bi-local Mapping}
\label{sec:bi-local mapping}
In the collective field description of the higher spin/vector model duality,
it was suggested that the bulk AdS higher spin theory is completely reconstructed by the bi-local collective field \cite{Das:2003vw},
which is in the canonical picture \cite{Koch:2014aqa} defined by 
	\begin{align}
		\Psi(t; \vec{x}_1, \vec{x}_2) \, \equiv \, \sum_{i=1}^N \phi_i(t, \vec{x}_1) \phi_i(t, \vec{x}_2) \, ,
	\end{align}
where $\phi_i(t, \vec{x})$ is the free $O(N)$ vector field.
In the light cone gauge, the bi-local map between the pure AdS higher spin canonical variables and the bi-local collective Minkowski CFT canonical variables 
was explicitly constructed in \cite{Koch:2010cy} by comparing the $SO(2, 3)$ global generators of the both theories.
Similar construction was given in \cite{Koch:2014aqa} for the time-like frame.

In the thermofield dynamics description of the AdS black hole higher spin theory, it was suggested in \cite{Jevicki:2015sla} to require for the gauge singlet state to be
	\begin{align}
		\big( J_{ij}^R + J_{ij}^L \big) | \Psi \rangle \, = \, 0 \, ,
	\end{align}
where $J_{ij}^{R(L)}$ is the $O(N)$ gauge symmetry generator for the right (left) boundary CFT vector model.
This requirement implies that we have three kinds of the bi-local collective oscillators
	\begin{align}
		\alpha_{RR}(\vec{p}_1, \vec{p}_2) \, =& \ \sum_{i=1}^N \, b_i^R(\vec{p}_1) b_i^R(\vec{p}_2) \, , \nonumber\\
		\alpha_{LL}(\vec{p}_1, \vec{p}_2) \, =& \ \sum_{i=1}^N \, b_i^L(\vec{p}_1) b_i^L(\vec{p}_2) \, , \nonumber\\
		\gamma_{RL}(\vec{p}_1, \vec{p}_2) \, =& \ \sum_{i=1}^N \, b_i^R(\vec{p}_1) b_i^L(\vec{p}_2) \, ,
	\label{bi-local oscillators}
	\end{align}
where $b_i^{R(L)}(\vec{p})$ are the oscillators of the right (left) boundary vector model.
In the previous study \cite{Jevicki:2015sla} it was demonstrated that the linearized Large $N$ dynamics for these three (coupled) bi-locals
leads to higher spin fields in asymptotically AdS spacetime. 
It was crucially noted that the mixed bi-local field $\gamma$ plays a central role in the construction: it secures the connectedness of the gravitational spacetime.

\subsection{Transformation from Poincare to Rindler-AdS Coordinates}
\label{sec:transformation from poincare to rindler-ads coordinates}
It is well known that AdS$_{d+1}$ space can be embedded into $(d+2)$-dimensional flat space.
For AdS$_4$ case, it is described by a hypersurface
	\begin{align}
		- (X^0)^2 \, + \, (X^1)^2 \, + \, (X^2)^2 \, + \, (X^3)^2 \, - \, (X^5)^2 \, = \, - 1 \, ,
	\end{align}
in $\mathbb{R}^{(2,3)}$ space.

The Poincare coordinates $(t, x, y, z)$ are achieved by the following parametrization of the embedding space:
	\begin{gather}
		X^0 \, = \, \frac{t}{z} \, , \qquad
		X^1 \, = \, \frac{y}{z} \, , \qquad
		X^2 \, = \, \frac{x}{z} \, , \nonumber\\
		X^3 \, = \, \frac{1}{2z} \Big[ \vec{x}\, {}^2 \, + \, z^2 \, - t^2 \, - \, 1 \Big] \, , \qquad
		X^5 \, = \, \frac{1}{2z} \Big[ \vec{x}\, {}^2 \, + \, z^2 \, - t^2 \, + \, 1 \Big] \, ,
	\label{Poincare-embedding}
	\end{gather}
where $\vec{x}=(x, y)$. This parametrization induces the Poincare metric
	\begin{align}
		ds^2 \, = \, \frac{1}{z^2} \Big[ \, - dt^2 \, + \, dx^2 \, + \, dy^2 \, + \, dz^2 \, \Big] \, .
	\end{align}

On the other hand, the right Rindler-${\rm AdS}$ wedge $(\tau, x, \sigma, \rho)$ is constructed by
	\begin{gather}
		X^0 \, = \, \frac{\sqrt{1-\rho^2}}{\rho} \, \sinh \tau \, , \qquad
		X^1 \, = \, \frac{\sqrt{1-\rho^2}}{\rho} \, \cosh \tau \, , \qquad
		X^2 \, = \, \frac{x}{\rho \, \sigma} \, , \nonumber\\
		X^3 \, = \, \frac{1}{2 \rho \sigma} \Big[ x^2 \, + \, \sigma^2 \, - \, 1 \Big] \, , \qquad
		X^5 \, = \, \frac{1}{2 \rho \sigma} \Big[ x^2 \, + \, \sigma^2 \, + \, 1 \Big] \, ,
	\label{Rindler-embedding}
	\end{gather}
which induces the Rindler-${\rm AdS}$ metric (\ref{Rindler-AdS}).

By equating (\ref{Poincare-embedding}) to (\ref{Rindler-embedding}), one can find explicit transformations from one to the other.
The transformations from the Poincare coordinates $(t, x, y, z)$ to the right Rindler-AdS coordinates $(\tau, x, \sigma, \rho)$ are
	\begin{align}
		\tau \, = \, \tanh^{-1}\left( \frac{t}{y} \right) , \quad
		x \, = \, x \, , \quad
		\sigma \, = \, \sqrt{-t^2+y^2+z^2} \, , \quad
		\rho \, = \, \frac{z}{\sqrt{-t^2+y^2+z^2}} \, .
	\end{align}
The transformations from the right Rindler-AdS coordinates to the Poincare coordinates are
 	\begin{align}
		t \, = \, \sigma \sqrt{1-\rho^2} \, \sinh \tau \, , \qquad
		x \, = \, x \, , \qquad
		y \, = \, \sigma \sqrt{1-\rho^2} \, \cosh \tau \, , \qquad
		z \, = \, \rho \, \sigma \, .
	\label{r-to-p-1}
	\end{align}

Now that we know the explicit transformations between the Poincare and the right Rindler coordinates, we can also write down the transformations in the momentum spaces.
In this paper, we use the first-quantized particle picture.
In that case, the momentum is just derivatives of its coordinates, so by chain rule, the Rindler momenta are related to the Poincare momenta.
Therefore, we find
	\begin{align}
		p^{\tau} \, =& \ y \, p^t \, - \, t \, p^y \, , \nonumber\\
		p^x \, =& \ p^x \, , \nonumber\\
		p^{\sigma} \, =& \ \frac{(- t \, p^t \, + \, y \, p^y \, + \, z \, p^z)}{\sqrt{ - t^2 + y^2 + z^2}} \, , \nonumber\\
		p^{\rho} \, =& \ \left[ \, \frac{z \, (t \, p^t - y \, p^y)}{-t^2+y^2} \, + \, p^z \right] \sqrt{-t^2+y^2+z^2} \, .
	\end{align}
The inverse transformations are also given by
	\begin{align}
		p^t \, =& \ \frac{1}{\sigma\sqrt{1- \rho^2}} \Big[ p^{\tau} \cosh \tau \, + \, (\sigma \, p^{\sigma} \, - \, \rho \, p^{\rho})(1-\rho^2) \sinh \tau \Big] \, , \nonumber\\
		p^x \, =& \ p^x \, , \nonumber\\
		p^y \, =& \ \frac{1}{\sigma\sqrt{1- \rho^2}} \Big[ p^{\tau} \sinh \tau \, + \, (\sigma \, p^{\sigma} \, - \, \rho \, p^{\rho})(1-\rho^2) \cosh \tau \Big] \, , \nonumber\\
		p^z \, =& \ \rho \, p^{\sigma} \, + \, \left( \frac{1-\rho^2}{\sigma} \right) p^{\rho} \, .
	\end{align}

From these results, one can check that the canonical commutation relations for the Rindler coordinates 
	\begin{align}
		- \, [\, \tau, \, p^{\tau}] \, = \, [\, x, \, p^{x}] \, = \, [\, \sigma, \, p^{\sigma}] \, = \, [\, \rho, \, p^{\rho}] \, = \, 1 \, ,
	\end{align}
are actually satisfied, provided that
	\begin{align}
		- \, [\, t, \, p^0] \, = \, [\, x, \, p^x] \, = \, [\, y, \, p^y] \, = \, [\, z, \, p^z] \, = \, 1 \, ,
	\end{align}
and vice versa.
We can also check that the remaining commutators vanish.
Therefore, these transformations between the Poincare and Rindler coordinates are indeed canonical transformations.

\subsection{$SO(2,3)$ Generators}
\label{sec:so(2,3) generators}
In \cite{Koch:2014aqa}, the AdS$_4$ global symmetry generators are given in terms of the Poincare coordinates and its momenta at $t=0$.
By using the coordinate transformations (\ref{r-to-p-1}), one can rewrite these generators in the right Rindler coordinates.
First note that $t=0$ corresponds to $\tau=0$. Therefore the transformations are now
 	\begin{align}	
		t \, = \, 0 \, , \quad
		x \, = \, x \, , \quad
		y \, = \, \sigma \sqrt{1-\rho^2}\, , \quad
		z \, = \, \rho \, \sigma \, ,
	\label{t0-coor}
	\end{align}
and
	\begin{align}	
		p^t \, = \, \frac{p^{\tau}}{\sigma\sqrt{1- \rho^2}} \, , \quad
		p^x \, = \, p^x \, , \quad
		p^y \, = \, \frac{\sqrt{1- \rho^2}}{\sigma} \, (\sigma \, p^{\sigma} \, - \, \rho \, p^{\rho}) \, , \quad
		p^z \, = \, \rho \, p^{\sigma} \, + \, \left( \frac{1-\rho^2}{\sigma} \right) p^{\rho} \, .
	\label{t0-mom}
	\end{align}
Hence the generators of the right wedge are 
	\begin{align}
		\hat{P}_{\rm Rads}^0 \, =& \ \sqrt{(p^x)^2 \, + \, (p^{\sigma})^2 \, + \, \left( \frac{1-\rho^2}{\sigma^2} \right) (p^{\rho})^2 } \, , \label{Gen-Rads0} \\
		\hat{P}_{\rm Rads}^1 \, =& \ p^x \, , \\
		\hat{P}_{\rm Rads}^2 \, =& \ \frac{\sqrt{1- \rho^2}}{\sigma} \, (\sigma \, p^{\sigma} \, - \, \rho \, p^{\rho}) \, , \\
		\hat{J}_{\rm Rads}^{01} \, =& \, - x \, \hat{P}^0_{\rm Rads}
		\, - \, \frac{\sqrt{1-\rho^2} \, (\sigma \, p^{\sigma} - \rho \, p^{\rho}) \, \big[ \rho \,\sigma \,p^{\sigma} + (1-\rho^2) p^{\rho}\big] \, p^{\theta}}
		{\sigma^2 (p^x)^2 + (1-\rho^2)(\sigma \, p^{\sigma} - \rho \, p^{\rho} )^2 } \, , \\
		\hat{J}_{\rm Rads}^{12} \, =& \ \frac{x \sqrt{1-\rho^2}}{\sigma} \, (\sigma \, p^{\sigma} \, - \, \rho \, p^{\rho} ) \, - \, \sigma \sqrt{1-\rho^2} \ p^x \, , \\
		\hat{J}_{\rm Rads}^{20} \, =& \, \sigma \sqrt{1-\rho^2} \, \hat{P}^0_{\rm Rads}
		\, - \, \frac{ \sigma \, p^x \, \big[ \rho \,\sigma \,p^{\sigma} + (1-\rho^2) p^{\rho}\big] \, p^{\theta}}
		{\sigma^2 (p^x)^2 + (1-\rho^2)(\sigma \, p^{\sigma} - \rho \, p^{\rho} )^2 } \, , \\
		\hat{D}_{\rm Rads} \, =& \ x \, p^x \, + \, \sigma \, p^{\sigma} \, , \\
		\hat{K}_{\rm Rads}^0 \, =& \, - \frac{1}{2} (x^2 + \sigma^2) \hat{P}^0_{\rm Rads}
		\, - \, \frac{ 2 \sigma \, \big[ \rho \,\sigma \,p^{\sigma} + (1-\rho^2) p^{\rho}\big] \, \hat{J}^{12}_{\rm Rads} \, p^{\theta} \, + \sigma^2 \, (p^{\theta})^2 \, \hat{P}^0_{\rm Rads}}
		{2 \big[ \sigma^2 (p^x)^2 + (1-\rho^2)(\sigma \, p^{\sigma} - \rho \, p^{\rho} )^2 \big]} \, , \\
		\hat{K}_{\rm Rads}^1 \, =& \, - \frac{1}{2} (x^2 + \sigma^2) p^x \, + \, x \, \hat{D}_{\rm Rads}
		\, + \, \frac{ 2 \sigma^2 \rho \sqrt{1-\rho^2} (\sigma \, p^{\sigma} - \rho \, p^{\rho}) \hat{P}^0_{\rm Rads} \, p^{\theta} \, + \sigma^2 \, p^x \, (p^{\theta})^2}
		{2 \big[ \sigma^2 (p^x)^2 + (1-\rho^2)(\sigma \, p^{\sigma} - \rho \, p^{\rho} )^2 \big]} \, , \\
		\hat{K}_{\rm Rads}^2 \, =& \, - \frac{1}{2\sigma} \sqrt{1-\rho^2} \, (x^2 + \sigma^2) (\sigma \, p^{\sigma} - \rho \, p^{\rho})
		\, + \, \sigma \sqrt{1-\rho^2} \, \hat{D}_{\rm Rads} \nonumber\\
		&\qquad - \, \frac{ 2 \sigma^3 \rho \, p^x \, \hat{P}_{\rm Rads}^0 \, p^{\theta} \, - \sigma \sqrt{1-\rho^2} \, (\sigma \, p^{\sigma} - \rho \, p^{\rho}) \, (p^{\theta})^2}
		{2 \big[ \sigma^2 (p^x)^2 + (1-\rho^2)(\sigma \, p^{\sigma} - \rho \, p^{\rho} )^2 \big]} \, . \label{Gen-Rads1}
	\end{align}
There generators also satisfy the $SO(2,3)$ algebra.
In the same way, one can also construct the generators in the left Rindler wedge.

Next, we would like to construct the $SO(2,3)$ bi-local conformal generators.
The $SO(2,3)$ generators of the conformal symmetry in 3d Minkowski space ($t,x,y$) are well known.
When $t=0$ ($\tau=0$), the transformations from 3d Minkowski to the right Minkowski-Rindler wedge are given by
 	\begin{align}
		x \, = \, x \, , \quad
		y \, = \, \sigma \, , \quad
		p^x \, = \, p^x \, , \quad
		p^y \, = \, p^{\sigma} \, .
	\label{coor-cft}
	\end{align}
This is just a trivial change of naming of the coordinates.
Anyway, now one can construct the generators in the right Minkowski-Rindler wedge from the Minkowski ones.
Then, one can easily construct the bi-local generators as
	\begin{align}
		\hat{p\, }_{\rm RRcft}^0 \, =& \ \sqrt{(p_1^x)^2 + (p_1^{\sigma})^2} \, + \, \sqrt{(p_2^x)^2 + (p_2^{\sigma})^2} \, , \label{Gen-RRcft0} \\
		\hat{p\, }_{\rm RRcft}^1 \, =& \ \, p_1^x \, + \, p_2^x \, , \\
		\hat{p\, }_{\rm RRcft}^2 \, =& \ \, p_1^{\sigma} \, + \, p_2^{\sigma} \, ,
	\end{align}
	\begin{align}
		\hat{j\, }_{\rm RRcft}^{01} \, =& \, - x_1 \, \sqrt{(p_1^x)^2 + (p_1^{\sigma})^2} \, - \, x_2 \, \sqrt{(p_2^x)^2 + (p_2^{\sigma})^2} \, , \\
		\hat{j\, }_{\rm RRcft}^{12} \, =& \ x_1 \, p_1^{\sigma} \, + \, x_2 \, p_2^{\sigma} \, - \, \sigma_1 \, p_1^x \, - \, \sigma_2 \, p_2^x \, , \\
		\hat{j\, }_{\rm RRcft}^{20} \, =& \ \sigma_1 \, \sqrt{(p_1^x)^2 + (p_1^{\sigma})^2} \, + \, \sigma_2 \, \sqrt{(p_2^x)^2 + (p_2^{\sigma})^2} \, , \\
		\hat{d\, }_{\rm RRcft} \, =& \ x_1 \, p_1^x \, + \, x_2 \, p_2^x \, + \, \sigma_1 \, p_1^{\sigma} \, + \, \sigma_2 \, p_2^{\sigma} \, , \\
		\hat{k\, }_{\rm RRcft}^0 \, =& \, - \frac{1}{2} (x_1^2 + \sigma_1^2) \, \sqrt{(p_1^x)^2 + (p_1^{\sigma})^2}
		\, - \, \frac{1}{2} (x_2^2 + \sigma_2^2) \, \sqrt{(p_2^x)^2 + (p_2^{\sigma})^2} \, , \\
		\hat{k\, }_{\rm RRcft}^1 \, =& \ x_1 (x_1 \, p_1^x \, + \, \sigma_1 \, p_1^{\sigma}) \, + \, x_2 (x_2 \, p_2^x \, + \, \sigma_2 \, p_2^{\sigma})
		\, - \, \frac{1}{2} (x_1^2 + \sigma_1^2) \, p_1^x \, - \, \frac{1}{2} (x_2^2 + \sigma_2^2) \, p_2^x \, , \\
		\hat{k\, }_{\rm RRcft}^2 \, =& \ \sigma_1 (x_1 \, p_1^x \, + \, \sigma_1 \, p_1^{\sigma}) \, + \, \sigma_2 (x_2 \, p_2^x \, + \, \sigma_2 \, p_2^{\sigma}) 
		\, - \, \frac{1}{2} (x_1^2 + \sigma_1^2) \, p_1^{\sigma} \, - \, \frac{1}{2} (x_2^2 + \sigma_2^2) \, p_2^{\sigma} \, .
	\label{Gen-RRcft1}
	\end{align}
The left generators are also constructed in the same way.

\subsection{Bi-local Map for the Rindler-AdS}
\label{sec:bi-local map for the rindler-ads}
In \cite{Koch:2014aqa}, the momentum space bi-local map from the Minkowski bi-local space ($t; x_1, y_1; x_2, y_2$) whose metric is
	\begin{align}
		ds^2 \, = \, - \, dt^2 \, + \, dx^2 \, + \, dy^2 \, ,
	\end{align}
to the ${\rm AdS_4}$ higher spin Poincare coordinate ($t, x, y, z; \theta$) with the metric
	\begin{align}
		ds^2 \, = \, \frac{1}{z^2} \Big[ - dt^2 \, + \, dx^2 \, + \, dy^2 \, + \, dz^2 \Big] \, ,
	\end{align}
is given by
	\begin{align}
		p^t \, =& \ \, |\vec{p}_1| \, + \, |\vec{p}_2| \, , \nonumber\\
		p^x \, =& \ \, p_1^x \, + \, p_2^x \, , \nonumber\\
		p^y \, =& \ \, p_1^y \, + \, p_2^y \, , \nonumber\\
		p^z \, =& \ \, \sqrt{2 \, |\vec{p}_1||\vec{p}_2| - 2 \, \vec{p}_1 \cdot \vec{p}_2 \, } \, , \nonumber\\
		\theta \ =& \ \, \arctan \left( \frac{2 \, (\, p_2^x \, p_1^y - p_2^y \, p_1^x \, )}{(|\vec{p}_2| - |\vec{p}_1|) \, p^z}\right) \, ,
	\end{align}
where $\vec{p}_1=(p_1^x, p_1^y)$, $\vec{p}_2=(p_2^x, p_2^y)$.
The induced coordinate transformations are
	\begin{align}
		x \, =& \ \frac{|\vec{p}_1|x_1+|\vec{p}_2|x_2}{\sqrt{(p^z)^2+(\vec{p})^2}} \, - \, \frac{p^y p^z p^{\theta}}{(\vec{p})^2\sqrt{(p^z)^2+(\vec{p})^2}} \, , \nonumber\\
		y \, =& \ \frac{|\vec{p}_1|y_1+|\vec{p}_2|y_2}{\sqrt{(p^z)^2+(\vec{p})^2}} \, + \, \frac{p^x p^z p^{\theta}}{(\vec{p})^2\sqrt{(p^z)^2+(\vec{p})^2}} \, , \nonumber\\
		z \, =& \ \frac{(\vec{x}_1-\vec{x}_2) \cdot (|\vec{p}_2|\vec{p}_1 - |\vec{p}_1| \vec{p}_2)}{p^z (|\vec{p}_1| + |\vec{p}_2|)} \, , \nonumber\\
		p^{\theta} \, =& \ (x_2- x_1) \sqrt{|\vec{p}_1||\vec{p}_2|} \cos \left( \frac{\varphi_1+\varphi_2}{2} \right) \nonumber\\
		&\, + \, (y_2- y_1) \sqrt{|\vec{p}_1||\vec{p}_2|} \sin \left( \frac{\varphi_1+\varphi_2}{2} \right) \, ,
	\end{align}
where $\vec{x}_1=(x_1, y_1)$, $\vec{x}_2=(x_2, y_2)$.

By using the transformations (\ref{t0-coor}) (\ref{t0-mom}) and (\ref{coor-cft}),
one can construct the bi-local maps from 3d bi-local regular Rindler space ($\tau; x_1, \sigma_1; x_2, \sigma_2$) whose metric is
	\begin{align}
		ds^2 \, = \, - \sigma^2 d\tau^2 \, + \, dx^2 \, + \, d\sigma^2 \, ,
	\end{align}
to the Rindler-AdS space ($\tau, x, \sigma, \rho; \theta$) with the metric
	\begin{align}
		ds^2 \, = \, \frac{1}{\rho^2} \left[ -(1-\rho^2) d\tau^2 \, + \, \frac{d\rho^2}{1-\rho^2} \, + \, \frac{dx^2 + d\sigma^2}{\sigma^2} \right] \, ,
	\end{align}
as
	\begin{align}
		p^{\tau} \, =& \ \, \sigma \sqrt{1-\rho^2} \, (|\vec{p}_1| \, + \, |\vec{p}_2| ) \, , \nonumber\\
		p^x \, =& \ \, p_1^x \, + \, p_2^x \, , \nonumber\\
		p^{\sigma} \, =& \ \, \rho \sqrt{ 2 |\vec{p}_1| |\vec{p}_2| - 2 \, \vec{p}_1 \cdot \vec{p}_2 \, } \, + \, \sqrt{1-\rho^2} \, (p^{\sigma}_1+p^{\sigma}_2) \, , \nonumber\\
		p^{\rho} \, =& \ \, \sigma \sqrt{ 2 |\vec{p}_1| |\vec{p}_2| - 2 \, \vec{p}_1 \cdot \vec{p}_2 \, }
		\, - \, \frac{\sigma \rho}{\sqrt{1-\rho^2}} \, (p_1^{\sigma}+p_2^{\sigma})\, , \nonumber\\
		\theta \ =& \ \, \arctan \left( \frac{2 \, (\, p_2^x \, p_1^{\sigma} - p_2^{\sigma} \, p_1^x \, )}
		{(|\vec{p}_2| - |\vec{p}_1|) \, \sqrt{ 2 |\vec{p}_1| |\vec{p}_2| - 2 \, \vec{p}_1 \cdot \vec{p}_2 \, } } \right) \, ,
	\label{map-mom}
	\end{align}
and
	\begin{align}
		x \, =& \ \frac{|\vec{p}_1| x_1 + |\vec{p}_2| x_2}{|\vec{p}_1|+|\vec{p}_2|}
		\, - \, \frac{(p_1^{\sigma}+p_2^{\sigma})\sqrt{ 2 |\vec{p}_1| |\vec{p}_2| - 2 \, \vec{p}_1 \cdot \vec{p}_2 } \ p^{\theta}}
		{(|\vec{p}_1|+|\vec{p}_2|) \, [(p_1^x+p_2^x)^2+(p_1^{\sigma}+p_2^{\sigma})^2]} \, , \nonumber\\\
		\sigma \, =& \ \frac{z}{\rho} \, , \nonumber\\
		\rho \, =& \ \bigg( 1 \, + \, \frac{y^2}{z^2} \bigg)^{-\frac{1}{2}} \, , \nonumber\\
		p^{\theta} \, =& \ (x_2 - x_1) \sqrt{|\vec{p}_1||\vec{p}_2|} \cos \left( \frac{\varphi_1+\varphi_2}{2} \right)
		\, + \, (\sigma_2- \sigma_1) \sqrt{|\vec{p}_1||\vec{p}_2|} \sin \left( \frac{\varphi_1+\varphi_2}{2} \right) \, ,
	\label{map-coor}
	\end{align}
where we have defined $\vec{x}_1=(x_1, \sigma_1)$, $\vec{x}_2=(x_2, \sigma_2)$,
$\vec{p}_1=(p_1^x, p_1^{\sigma})$, $\vec{p}_2=(p_2^x, p_2^{\sigma})$,
$\varphi_1=\arctan(\frac{p_1^{\sigma}}{p_1^x})$, $\varphi_2=\arctan(\frac{p_2^{\sigma}}{p_2^x})$,
and $y$ and $z$ are functions of the bi-local regular Rindler canonical variables
	\begin{align}
		y \, =& \ \frac{|\vec{p}_1| \sigma_1 + |\vec{p}_2| \sigma_2}{|\vec{p}_1|+|\vec{p}_2|}
		\, - \, \frac{(p_1^x+p_2^x)\sqrt{ 2 |\vec{p}_1| |\vec{p}_2| - 2 \, \vec{p}_1 \cdot \vec{p}_2 } \ p^{\theta}}
		{(|\vec{p}_1|+|\vec{p}_2|) \, [(p_1^x+p_2^x)^2+(p_1^{\sigma}+p_2^{\sigma})^2]} \, , \nonumber\\
		z \, =& \ \frac{(\vec{x}_1-\vec{x}_2) \cdot (|\vec{p}_2|\vec{p}_1 - |\vec{p}_1| \vec{p}_2)}
		{(|\vec{p}_1| + |\vec{p}_2|) \sqrt{ 2 |\vec{p}_1| |\vec{p}_2| - 2 \, \vec{p}_1 \cdot \vec{p}_2 \, }} \, .
	\label{yz}
	\end{align}

The Rindler-AdS momenta induced by the bi-local map (\ref{map-mom}) indeed satisfy the Rindler-AdS on-shell condition
	\begin{align}
		0 \, = \, g^{\mu\nu} \partial_{\mu} \partial_{\nu}
		\, = \, - \, \frac{(p^{\tau})^2}{\sigma^2 (1-\rho^2)} \, + \, \left( \frac{1-\rho^2}{\sigma^2} \right) (p^{\rho})^2 \, + \, (p^x)^2 \, + \, (p^{\sigma})^2 \, .
	\end{align}
One can also check that the bi-local map (\ref{map-mom}), (\ref{map-coor}) actually maps the right Rindler-AdS generators (\ref{Gen-Rads0}) - (\ref{Gen-Rads1})
to the RR bi-local CFT generators (\ref{Gen-RRcft0}) - (\ref{Gen-RRcft1}).
Furthermore, the induced Rindler-AdS canonical valuables satisfy the canonical commutation relations
	\begin{align}
		[x \, , \, p^x ] \, = \, [\sigma \, , \, p^{\sigma}] \, = \, [\rho \, , \, p^{\rho}] \, = \, [\theta \, , \, p^{\theta}] \, = \, 1 \, ,
	\end{align}
with the other commutators vanishing, as a consequence of the bi-local commutator relations
	\begin{align}
		[x_1 \, , \, p^x_1 ] \, = \, [x_2 \, , \, p^x_2 ] \, = \, [\sigma_1 \, , \, p^{\sigma}_1] \, = \, [\sigma_2 \, , \, p^{\sigma}_2] \, = \, 1 \, ,
	\end{align}
and others vanish.

Now let us consider the bi-local oscillators defined in (\ref{bi-local oscillators}).
The diagonal oscillators $\alpha_{RR}^{\dagger}$, $\alpha_{LL}^{\dagger}$ create states with the following momenta
 	\begin{align}
		\alpha_{RR}^{\dagger}(\vec{p}_1, \vec{p}_2) \, : \quad p^{\tau} \, =& \ \sigma \sqrt{1-\rho^2} \Big( |\vec{p}_1| \, + \, |\vec{p}_2| \Big) \, , \nonumber\\
		p^x \, =& \ p^x_1 \, + \, p^x_2 \, , \nonumber\\
		p^{\sigma} \, =& \ \rho \sqrt{2 \, |\vec{p}_1| |\vec{p}_2| - 2 \, \vec{p}_1 \cdot \vec{p}_2 \, } \, + \, \sqrt{1-\rho^2} \, (p^{\sigma}_1 + p^{\sigma}_2) \, , \nonumber\\
		p^{\rho} \, =& \ \sigma \sqrt{2 \, |\vec{p}_1| |\vec{p}_2| - 2 \, \vec{p}_1 \cdot \vec{p}_2 \, } \, - \, \frac{\sigma\rho}{\sqrt{1-\rho^2}} (p^{\sigma}_1 + p^{\sigma}_2) \nonumber\\
		=& \ 2 \sigma \sqrt{|\vec{p}_1| |\vec{p}_2|} \, \sin \left( \frac{\varphi_1 - \varphi_2}{2} \right) \, - \, \frac{\sigma\rho}{\sqrt{1-\rho^2}} (p^{\sigma}_1 + p^{\sigma}_2) \, ,
	\end{align}
and
 	\begin{align}
		\alpha_{LL}^{\dagger}(\vec{p}_1, \vec{p}_2) \, : \quad p^{\tau} \, =& \ - \, \sigma \sqrt{1-\rho^2} \Big( |\vec{p}_1| \, + \, |\vec{p}_2| \Big) \, , \nonumber\\
		p^x \, =& \ - \, p^x_1 \, - \, p^x_2 \, , \nonumber\\
		p^{\sigma} \, =& \ \rho \sqrt{2 \, |\vec{p}_1| |\vec{p}_2| - 2 \, \vec{p}_1 \cdot \vec{p}_2 \, } \, - \, \sqrt{1-\rho^2} \, (p^{\sigma}_1 + p^{\sigma}_2) \, , \nonumber\\
		p^{\rho} \, =& \ \sigma \sqrt{2 \, |\vec{p}_1| |\vec{p}_2| - 2 \, \vec{p}_1 \cdot \vec{p}_2 \, } \, + \, \frac{\sigma\rho}{\sqrt{1-\rho^2}} (p^{\sigma}_1 + p^{\sigma}_2) \nonumber\\
		=& \ 2 \sigma \sqrt{|\vec{p}_1| |\vec{p}_2|} \, \sin \left( \frac{\varphi_1 - \varphi_2}{2} \right) \, + \, \frac{\sigma\rho}{\sqrt{1-\rho^2}} (p^{\sigma}_1 + p^{\sigma}_2) \, .
	\end{align} 
For these diagonal oscillators, the both terms on the right-hand side of $p^{\rho}$ are real.
Therefore, these oscillators create only propagating modes,
and cannot create evanescent modes, which characterized by an imaginary radial-direction momentum, anywhere in bulk Rindler-AdS space.
 
Next, we consider the off-diagonal oscillators $\gamma_{RL}^{\dagger}$.
This oscillator creates states with
 	\begin{align}
		\gamma_{RL}^{\dagger}(\vec{p}_1, \vec{p}_2) \, : \quad p^{\tau} \, =& \ \sigma \sqrt{1-\rho^2} \Big( |\vec{p}_1| \, - \, |\vec{p}_2| \Big) \, , \nonumber\\
		p^x \, =& \ p^x_1 \, - \, p^x_2 \, , \nonumber\\
		p^{\sigma} \, =& \ \rho \sqrt{2 \, \vec{p}_1 \cdot \vec{p}_2 - 2 \, |\vec{p}_1| |\vec{p}_2| \, } \, + \, \sqrt{1-\rho^2} \, (p^{\sigma}_1 - p^{\sigma}_2) \, , \nonumber\\
		p^{\rho} \, =& \ \sigma \sqrt{2 \, \vec{p}_1 \cdot \vec{p}_2 - 2 \, |\vec{p}_1| |\vec{p}_2| \, } \, - \, \frac{\sigma\rho}{\sqrt{1-\rho^2}} (p^{\sigma}_1 - p^{\sigma}_2) \nonumber\\
		=& \ 2 i \sigma \sqrt{|\vec{p}_1| |\vec{p}_2|} \, \sin \left( \frac{\varphi_1 - \varphi_2}{2} \right) \, - \, \frac{\sigma\rho}{\sqrt{1-\rho^2}} (p^{\sigma}_1 - p^{\sigma}_2) \, .
	\end{align}
Here, in the right-hand side of $p^{\rho}$, the first term is imaginary and the second term takes in general complex value.
Thus, this state has exponentially growing amplitude, which we can identify as an evanescent mode.

Next we discuss ``quasinormal modes'', which represents energy dissipation of a perturbed event horizon.
In \cite{Aros:2002}, the authors showed that in four and higher dimensional Rindler-AdS space (massless topological black hole),
quasinormal modes exist and the solutions can be exactly obtainable. 
The quasinormal frequency takes complex value. In fact, we can see the bi-local map is able to produce complex valued frequencies as we will see below.
The frequency $\omega$ is a conjugate valuable of $\tau$; therefore, $\omega=p^{\tau}$.
Rearranging the bi-local map for $p^{\tau}$, one can obtain
 	\begin{align}
		\alpha_{RR}^{\dagger}(\vec{p}_1, \vec{p}_2) \, : \quad \omega \, =& \ y \, \Big( |\vec{p}_1| \, + \, |\vec{p}_2| \Big) \, , \\
		\alpha_{LL}^{\dagger}(\vec{p}_1, \vec{p}_2) \, : \quad \omega \, =& \ - y \, \Big( |\vec{p}_1| \, + \, |\vec{p}_2| \Big) \, , \\
		\gamma_{RL}^{\dagger}(\vec{p}_1, \vec{p}_2) \, : \quad \omega \, =& \ y \, \Big( |\vec{p}_1| \, - \, |\vec{p}_2| \Big) \, ,
	\end{align}
where $y$ is a bi-local parameter which is given in (\ref{yz}).
For $\alpha_{RR}^{\dagger}$ and $\alpha_{LL}^{\dagger}$, the parameter $y$ is always real;
hence the frequency produced by $\alpha_{RR}^{\dagger}$ or $\alpha_{LL}^{\dagger}$ is always real.
On the other hand, for $\gamma_{RL}^{\dagger}$,
the second term of the right-hand side of $y$ is imaginary; therefore, along with the first term, the parameter $y$ can take a complex value.
Thus, we conclude that $\gamma_{RL}^{\dagger}$ is able to produce complex valued frequencies which contain quasinormal modes,
while $\alpha_{RR}^{\dagger}$ and $\alpha_{LL}^{\dagger}$ produce only real valued frequencies corresponding to propagating modes.
The origin of this complex frequency is obviously the coordinate dependence part of the bi-local map for $\omega$.
This coordinate mixing does not happen in the pure AdS case \cite{Koch:2010cy, Koch:2014aqa}; therefore, we obtain only real valued frequencies for the pure AdS background.
The coordinate dependence part ($\sqrt{1-\rho^2}$) of the bi-local map for $\omega$ implies the presence of the event horizon ($\rho=1$);
this agrees with the physical meaning of the quasinormal modes.

Finally, we also note that, if we have only the diagonal oscillators $\alpha_{RR}^{\dagger}$, $\alpha_{LL}^{\dagger}$,
the bi-loacl map for the Rindler-AdS coordinates (\ref{map-coor}) can produce only outside of the horizon $0<\rho<1$.
However, the entangled oscillator $\gamma_{RL}^{\dagger}$ can naturally produce the inside of the horizon $\rho>1$,
because for the mode created by this oscillator, the parameter $z$ has a pure imaginary value.

\section{Conclusion}
\label{sec:conclusion}
In this work we have presented a further study of the Thermo-field CFT duality in the case of $O(N)$ vector field theories.
It was shown that the thermofield system provides the degrees of freedom for a complete reconstruction of all-spin fields in the gravitational Rindler-AdS spacetime background.
The stability of the $O(N)$ vector model CFT on hyperbolic space is clarified.
Most significantly we have proved a map from a set of bi-locals to the connected Rindler-AdS spacetime.
This specific construction also provides implications of the very interesting question of reconstructing bulk observables behind the horizon.
A central role is played in the mixed set of bi-local observables representing singlet contractions between the left and right CFT's.
Their inclusion is natural in the diagonal $O(N)$ singlet collective description, but we note that they would not be included in the standard ``doubling'' of single trace CFT observables.

\acknowledgments
We are grateful to A.~Maloney for some clarification.
We would also like to thank S.~G.~Avery and J.~Yoon for helpful discussions.
This work is supported by the Department of Energy under contract DE-FG02-91ER40688.

\appendix
\section{Evanescent Modes in Rindler-AdS}
\label{sec:evanescent modes in rindler-ads}
Even though the existence of the evanescent modes in Rindler-AdS background was suggested in the discussion of smearing function \cite{Hamilton:2006, Bousso:2012mh},
in this section, we demonstrate the evanescent modes in this background from the point of view of the effective potential as in \cite{Rey:2014dpa}.

The equation of motion for a minimally coupled scalar field
	\begin{align}
		\partial_{\mu} \left( \sqrt{-g} \, g^{\mu\nu} \partial_{\nu} \Phi \right) \, = \, \sqrt{-g} \, m^2 \Phi \, ,
	\end{align}
for this background (\ref{Rindler-AdS}) is explicitly given by 
	\begin{align}
		- \, \frac{1}{1-\rho^2} \, \partial_{\tau}^2 \Phi \, + \, \rho^2 \partial_{\rho} \left( \frac{1-\rho^2}{\rho^2} \, \partial_{\rho} \Phi \right)
		\, + \, \sigma^2 \Big( \partial_x^2 \, \Phi \, + \, \partial_{\sigma}^2 \, \Phi \Big) \, = \, \frac{m^2}{\rho^2} \, \Phi \, .
	\end{align}
The solution for the $\tau$ and $x$ directions are plane waves, so we use 
	\begin{align}
		\Phi(\tau, x, \sigma, \rho) \, = \, e^{-i \omega \tau - i k_x x} \Phi(\sigma, \rho) \, .
	\end{align}
Then, the equation of motion for $\Phi(\sigma, \rho)$ is 
	\begin{align}
		\frac{\omega^2}{1-\rho^2} \, \Phi \, + \, \rho^2 \partial_{\rho} \left( \frac{1-\rho^2}{\rho^2} \, \partial_{\rho} \Phi \right)
		\, + \, \sigma^2 \Big( \partial_{\sigma}^2 \, - \, k_x^2 \Big) \Phi \, = \, \frac{m^2}{\rho^2} \, \Phi \, .
	\end{align}
The $\sigma$-direction solution is given by the modified Bessel function of the second kind as
	\begin{align}
		\Phi(\sigma, \rho) \, = \, \sigma^{\frac{1}{2}} K_{i \nu}(|k_x|\sigma) \ \Phi(\rho) \, ,
	\end{align}
which satisfies
	\begin{align}
		\sigma^2 \Big( \partial_{\sigma}^2 \, - \, k_x^2 \Big) \Phi(\sigma,z) \, = \, - \, \bar{\nu}^2 \, \Phi(\sigma, \rho) \, ,
	\end{align}
where $\nu$ is a free parameter which takes a real value, and $\bar{\nu}=\sqrt{\frac{1}{4}+\nu^2}$.
Thus, finally one obtains the $\rho$-direction equation of motion
	\begin{align}
		\left( \frac{\omega^2}{1-\rho^2} \, - \, \bar{\nu}^2 \, - \, \frac{m^2}{\rho^2} \right) \Phi(\rho)
		\, + \, \rho^2 \partial_{\rho} \left( \frac{1-\rho^2}{\rho^2} \, \partial_{\rho} \Phi(\rho) \right) \, = \, 0 \, .
	\label{rho-eq}
	\end{align}

\begin{figure}[t!]
	\begin{center}
		\scalebox{0.5}{\includegraphics{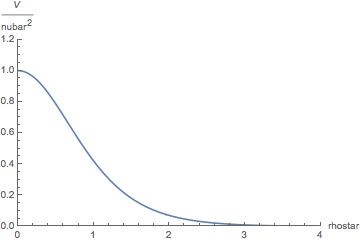}}
	\end{center}
	\caption{The ratio of the effective potential for Rindler-AdS, $V(\rho_*)$ with $\bar{\nu}^2$.
	On the boundary ($\rho_*=0$), $\omega^2\ge \bar{\nu}^2$; therefore, only propagating modes are permitted.
	On the other hand, near the horizon region ($\rho_*\to \infty$), the effective potential becomes zero.
	Accordingly, evanescent modes ($\omega^2 \sim 0$) are allowed to exist.}
\label{effective-potential}
\end{figure}

By excluding the boundary behavior of the field $\Phi(\rho)=\rho \, u(\rho)$, the radical direction equation is rewritten as 
	\begin{align}
		(1 - \rho^2) \, \partial_{\rho}^2 \, u \, - \, 2 \rho \, \partial_{\rho} \, u \, + \, \left( \frac{\omega^2}{1-\rho^2} \, - \, \bar{\nu}^2 \right) u \, = \, 0 \, ,
	\end{align}
where we have used $m^2=-2$. Moving to the tortoise coordinate $d\rho_*=\frac{d\rho}{(1-\rho^2)}$, (i.e. $\rho_*=\frac{1}{2} \log\left( \frac{1+\rho}{1-\rho} \right)$), 
we can obtain a ${\rm Schr\ddot{o}dinger}$-like equation,
	\begin{align}
		\frac{d^2u(\rho_*)}{d\rho_*^2} \, + \, \Big( \omega^2 - V(\rho_*) \Big) \, u(\rho_*) \, = \, 0 \, ,
 	\label{Schro-eq}
	\end{align}
where the effective potential $V(\rho_*)$ is given by
	\begin{align}
		V(\rho_*) \, = \, \frac{\bar{\nu}^2}{\cosh^2(\rho_*)} \, .
 	\end{align}
In the tortoise coordinate $\rho_*$, the AdS boundary is located at $\rho_*=0$, while the Rindler-horizon is at $\rho_*=\infty$.
The effective potential becomes zero near horizon $\rho_*\to\infty$ (see Figure~\ref{effective-potential}).
This potential shape is very similar to the case of a large black hole discussed in \cite{Rey:2014dpa}.
Therefore, such modes with $\omega=0$ are allowed to exist in the near horizon region and these modes are indeed evanescent modes.
One can verify this statement immediately seeing the WKB approximation solution of the equation (\ref{Schro-eq})
	\begin{align}
		u(\rho_*) \, \sim \, \exp \left( \, \int_{\rho_*}^{\infty} d\rho_*' \, \sqrt{V(\rho_*') - \omega^2} \, \right) \, .
	\end{align}
We are interested in the most evanescent mode $\omega^2=0$.
In this case, the integrand is real function; therefore, the solution is exponentially growing in the radical direction.
Actually, using the original coordinate $\rho$ and integrating the exponent, one finds $u(\rho) \sim \exp[\bar{\nu} \sin^{-1}(\rho)]$.
Hence, this is evanescent mode.

\section{Canonical Quantization of Rindler $O(N)$ Vector Model}
\label{sec:canonical quantization of rindler o(n) vector model}
In this appendix, we give a detail discussion of the canonical quantization for the $O(N)$ vector model in the Rindler spacetime.

From the Rindler action (\ref{action-rindler}), the Klein-Gordon equation of motion explicitly reads
	\begin{align}
		- \, (\partial_{\tau}^2 \phi_i^R) \, + \, \sigma \partial_{\sigma} \phi_i^R \, + \, \sigma^2 \partial_{\sigma}^2 \, \phi_i^R \, + \, \sigma^2 \partial_x^2 \, \phi_i^R \, = \, 0 \, .
	\end{align}
Since the solution for the $\tau$ and $x$ directions are just place-wave, we employ the following ansatz for the solution of the equation.
	\begin{align}
		g^R_{\omega, k}(\tau, x, \sigma) \, = \, \frac{1}{2\pi \sqrt{2\omega}} \, v^R(\omega, k, \sigma) \, e^{i k x - i \omega \tau } \, .
	\end{align}
Plugging this ansatz into the equation, we obtain
	\begin{align}
		\partial_{\tilde{\sigma}}^2 \, v^R \, + \, \frac{1}{\tilde{\sigma}} \, \partial_{\tilde{\sigma}} \, v^R \, + \, \left( \frac{\omega^2}{\tilde{\sigma}^2} - 1 \right) v^R \, = \, 0 \, , 
	\end{align}
where $\tilde{\sigma}= |k| \sigma$.

This equation is the modified Bessel equation and the solution is given by the modified Bessel function:
	\begin{align}
		v^R(\omega, k, \sigma) \, \propto \, K_{i \omega}(|k| \sigma) \, ,
	\end{align}
which is defined by 
	\begin{align}
		K_{i \omega}(x) \, = \, \frac{\pi}{2 \sinh(\pi \omega)} \Big[ I_{i\omega}(x) \, - \, I_{-i\omega}(x) \Big] \, ,
	\end{align}
and
	\begin{align}
		I_{i \omega}(x) \, = \, \sum_{n=0}^{\infty} \, \frac{1}{n! \, \Gamma(i\omega+n+1)} \left( \frac{x}{2} \right)^{2n+i\omega} \, .
	\end{align}
Hence, with an appropriate normalization, the general solution is
	\begin{align}
		\phi_i^R(\tau, x, \sigma) \, = \, \int_0^{\infty} d\omega \int_{-\infty}^{\infty} dk \Big[ \ b_i^R(\omega, k) \, g^R_{\omega, k}(\tau, x, \sigma)
		\, + \, b_i^{R\dagger}(\omega, k) \, g^{R*}_{\omega, k}(\tau, x, \sigma) \ \Big] \, ,
	\end{align}
where
	\begin{align}
		g^R_{\omega, k}(\tau, x, \sigma) \, = \, \frac{1}{2\pi \sqrt{2\omega}} \, \frac{2K_{i \omega}(|k| \sigma)}{|\Gamma(i\omega)|} \, e^{i k x - i \omega \tau} \, .
	\label{solution-g}
	\end{align}
The normalization of the solution is adjusted by using the Klein-Gordon inner product, which is defined by
	\begin{align}
		(\phi_1, \phi_2)_{\rm KG} \, = \, - i \int_{\Sigma} d\sigma dx \, \frac{\sqrt{-g}}{g_{00}} \, 
		\Big[ \, \phi_1^* \partial_{\tau} \phi_2 \, - \, \phi_2 \partial_{\tau} \phi_1^* \, \Big] \, ,
	\end{align}
where $\Sigma$ is a constant $\tau$ hypersurface.
This definition has the following identity.
	\begin{align}
		(\phi_2 \, , \, \phi_1)_{\rm KG} \, = \, - \, (\phi_1^* \, , \, \phi_2^*)_{\rm KG} \, = \, (\phi_1 \, , \, \phi_2)^*_{\rm KG} \, .
	\end{align}
First, note that the orthonormal relation of the modified Bessel function is given by \cite{Longhi}
	\begin{align}
		\int_0^{\infty} \frac{d\sigma}{\sigma} \, K_{i \omega}(\sigma) \, K_{-i\omega'}(\sigma) \, = \, \frac{\pi}{2} \, |\Gamma(i \omega)|^2 \, \delta(\omega - \omega') \, .
	\label{Bessel-orthogo}
	\end{align}
By using this, one can compute
	\begin{align}
		\int_{\Sigma} d\sigma dx \, \frac{\sqrt{-g}}{g_{00}} \, \Big[ \, g^{R}_{\omega, k}(\tau, \sigma, x) \ g^{R*}_{\omega', k'}(\tau, \sigma, x) \, \Big]
		\, = \, - \frac{1}{2\omega} \, \delta(\omega - \omega') \delta(k - k') \, .
	\end{align}
Thus, the solution is orthonormalized in terms of the Klein-Gordon inner product as
	\begin{align}
		(g^R_{\omega, k}, \, g^R_{\omega', k'})_{\rm KG} \, =& \ \delta(\omega - \omega') \delta(k - k') \, , \\
		(g^{R*}_{\omega, k}, \, g^R_{\omega', k'})_{\rm KG} \, =& \ 0 \, .
	\end{align}
Therefore, now the oscillators are written in terms of the Klein-Gordon inner product as
	\begin{align}
		b_i^R(\omega, k) \, =& \ (g^R_{\omega, k} , \phi_i)_{\rm KG} \nonumber\\
		=& \ \int d\sigma dx \Big[ \, i \, g^{R*}_{\omega, k}(\tau, \sigma, x) \, \Pi_i(\tau, \sigma, x)
		\, + \, \frac{\omega}{\sigma} \, \phi_i(\tau, \sigma, x) \, g^{R*}_{\omega, k}(\tau, \sigma, x) \, \Big] \, ,
	\end{align}
where $\Pi_i$ is the conjugate momentum defined by $\Pi_i=\sigma^{-1} \partial_{\tau} \phi_i$. Also one finds
	\begin{align}
		b_i^{R\dagger}(\omega , k) \, =& \ (\phi_i \, , \, g^R_{\omega, k})_{\rm KG} \nonumber\\
		=& \ \int d\sigma dx \Big[ \, \frac{\omega}{\sigma} \, \phi_i(\tau, \sigma, x) \, g^R_{\omega, k}(\tau, \sigma, x)
		\, - \, i \, g^R_{\omega, k}(\tau, \sigma, x) \, \Pi_i(\tau, \sigma, x) \, \Big] \, ,
	\end{align}
From these expressions, one can compute the commutation relations between the modes by requiring the canonical commutation relation of the field,
$\big[\phi_i(\tau, \sigma, x) \, , \, \Pi_j(\tau, \sigma, x)\big] = i \, \delta_{ij} \, \delta(\sigma-\sigma') \delta(x - x')$ as
	\begin{align}
		\big[ \, b_i^R(\omega, k) \, , \, b_j^{R\dagger}(\omega' , k') \, \big] \, = \, \delta_{ij} \delta(\omega- \omega') \, \delta(k - k') \, .
	\label{commutation-rindler}
	\end{align}

\section{Hamiltonian of Rindler Vector Model}
\label{sec:hamiltonian of rindler vector model}
In this appendix, we give a derivation for the Hamiltonian (\ref{hamiltonian}).
Here, we take $\tau=0$ hypersurface for $\Sigma$.
The kinetic term is computed as follows.
	\begin{align}
		\int_{\tau=0} d\sigma dx \, \sigma \, \Pi_i^2
		\, =& \ \int_{\tau=0} \frac{d\sigma}{\sigma} \, \int d\omega d\omega' dk dk' \frac{\sqrt{\omega \omega'}}{\pi} \nonumber\\
		& \times \bigg[ - b_i^R(\omega, k) \, b_i^R(\omega', k') \, \frac{K_{i\omega}(|k|\sigma) \, K_{i\omega'}(|k'|\sigma)}{|\Gamma(i\omega)| \, |\Gamma(i\omega')|} \,
		\delta(k+k') \nonumber\\
		& \hspace{18pt} + b_i^R(\omega, k) \, b_i^{R\dagger}(\omega', k') \,
		\frac{K_{i\omega}(|k|\sigma) \, K_{-i\omega'}(|k'|\sigma)}{|\Gamma(i\omega)| \, |\Gamma(-i\omega')|} \, \delta(k-k') \nonumber\\
		& \hspace{18pt} + b_i^{R\dagger}(\omega, k) \, b_i^R(\omega', k') \,
		\frac{K_{-i\omega}(|k|\sigma) \, K_{i\omega'}(|k'|\sigma)}{|\Gamma(-i\omega)| \, |\Gamma(i\omega')|} \, \delta(k-k') \nonumber\\
		& \hspace{18pt} - b_i^{R\dagger}(\omega, k) \, b_i^{R\dagger}(\omega', k') \,
		\frac{K_{-i\omega}(|k|\sigma) \, K_{-i\omega'}(|k'|\sigma)}{|\Gamma(-i\omega)| \, |\Gamma(-i\omega')|} \, \delta(k+k') \ \bigg] \nonumber\\
		=& \ \int d\omega dk \, \frac{\omega}{2} \bigg[ b_i^R(\omega, k) \, b_i^{R\dagger}(\omega, k) \, + \, b_i^{R\dagger}(\omega, k) \, b_i^R(\omega, k) \ \bigg] \, ,
	\end{align}
where we have used the orthonormal relation of the modified Bessel function (\ref{Bessel-orthogo}), and $\delta(\omega+\omega')=0$ because $\omega, \omega'>0$.

By similar ways, one can also find
	\begin{align}
		\int_{\tau=0} d\sigma dx \, \sigma \, (\partial_x \phi_i)^2
		\, =& \ \int \frac{d\omega d\omega' dk}{\pi \sqrt{\omega \omega'}} 
		\bigg[ - \frac{F_2(\omega, \omega')}{|\Gamma(i\omega)| \, |\Gamma(i\omega')|} \ b_i^R(\omega, k) \, b_i^R(\omega', -k) \nonumber\\
		& \hspace{68pt} + \frac{F_2(\omega, -\omega')}{|\Gamma(i\omega)| \, |\Gamma(-i\omega')|} \
		b_i^R(\omega, k) \, b_i^{R\dagger}(\omega', k) \nonumber\\
		& \hspace{68pt} + \frac{F_2(-\omega, \omega')}{|\Gamma(-i\omega)| \, |\Gamma(i\omega')|} \
		b_i^{R\dagger}(\omega, k) \, b_i^R(\omega', k) \nonumber\\
		& \hspace{68pt} - \frac{F_2(-\omega, -\omega')}{|\Gamma(-i\omega)| \, |\Gamma(-i\omega')|} \
		b_i^{R\dagger}(\omega, k) \, b_i^{R\dagger}(\omega', -k) \ \bigg] \, ,
	\end{align}
and
	\begin{align}
		\int_{\tau=0} d\sigma dx \, \sigma \, (\partial_{\sigma} \phi)^2
		\, =& \ \int \frac{d\omega d\omega' dk}{\pi \sqrt{\omega \omega'}} 
		\bigg[ \, \ \frac{F_3(\omega, \omega')}{|\Gamma(i\omega)| \, |\Gamma(i\omega')|} \ b_i^R(\omega, k) \, b_i^R(\omega', -k) \nonumber\\
		& \hspace{61pt} + \frac{F_3(\omega, -\omega')}{|\Gamma(i\omega)| \, |\Gamma(-i\omega')|} \ b_i^R(\omega, k) \, b_i^{R\dagger}(\omega', k) \nonumber\\
		& \hspace{61pt} + \frac{F_3(-\omega, \omega')}{|\Gamma(-i\omega)| \, |\Gamma(i\omega')|} \ b_i^{R\dagger}(\omega, k) \, b_i^R(\omega', k) \nonumber\\
		& \hspace{61pt} + \frac{F_3(-\omega, -\omega')}{|\Gamma(-i\omega)| \, |\Gamma(-i\omega')|} \ b_i^{R\dagger}(\omega, k) \, b_i^{R\dagger}(\omega', -k) \ \bigg] \, ,
	\end{align}
where
	\begin{align}
		F_2(\omega, \omega') \, =& \ \int_0^{\infty} d\sigma \, \sigma \, K_{i \omega}(\sigma) \, K_{i\omega'}(\sigma) \, , \nonumber\\
		F_3(\omega, \omega') \, =& \ \int_0^{\infty} d\sigma \, \sigma \, \Big( \partial_{\sigma} K_{i \omega}(\sigma) \Big) \, \Big( \partial_{\sigma}K_{i\omega'}(\sigma) \Big) \, .
	\end{align}

Explicit computation of $F_2, F_3$ is given in the following.
There is a nice formula about orthogonality relation of the modified Bessel function $K_{\nu}(x)$ \cite{Jeffrey}
	\begin{align}
		\int_0^{\infty} dx \, x^{-\lambda} K_{\mu}(ax) K_{\nu}(bx) \, =& \ \frac{2^{-2-\lambda}a^{-\nu+\lambda-1}b^{\nu}}{\Gamma(1-\lambda)}
		\Gamma\left( \frac{1-\lambda+\mu+\nu}{2} \right) \Gamma\left( \frac{1-\lambda-\mu+\nu}{2} \right) \nonumber\\
		& \ \times \Gamma\left( \frac{1-\lambda+\mu-\nu}{2} \right) \Gamma\left( \frac{1-\lambda-\mu-\nu}{2} \right) \nonumber\\
		& \ \times F\left( \frac{1-\lambda+\mu+\nu}{2} , \frac{1-\lambda-\mu+\nu}{2} ; 1-\lambda ; 1 - \frac{b^2}{a^2} \right) \, . \nonumber\\
		{\rm for} \ \ {\rm Re}(a+b)>0, \quad & {\rm Re}(\lambda) < 1 - |{\rm Re} (\mu)| - |{\rm Re} (\nu)| \, .
	\end{align}

By using this formula, one can compute
	\begin{align}
		F_2(\omega, \omega') \, =& \ \frac{1}{2} \ \Gamma\Big( 1+\frac{i}{2} (\omega+\omega') \Big) \Gamma\Big( 1+\frac{i}{2} (\omega-\omega') \Big)
		\Gamma\Big( 1-\frac{i}{2} (\omega-\omega') \Big) \Gamma\Big( 1-\frac{i}{2} (\omega+\omega') \Big) \nonumber\\
		=& \ \frac{\pi^2 (\omega^2 - \omega'^2)}{4[\cosh(\pi \omega) - \cosh(\pi \omega')]} \, .
	\end{align}
Also 
	\begin{align}
		F_3(\omega, \omega') \, =& \ \frac{1}{8} \, \bigg[\, \Gamma\Big( \frac{i\omega+i\omega'}{2} \Big) \, \Gamma\Big( \frac{2+i\omega-i\omega'}{2} \Big) \,
		\Gamma\Big( \frac{2-i\omega+i\omega'}{2} \Big) \, \Gamma\Big( \frac{4-i\omega-i\omega'}{2} \Big) \nonumber\\
		& \quad + \, \Gamma\Big( \frac{2+i\omega+i\omega'}{2} \Big) \, \Gamma\Big( \frac{i\omega-i\omega'}{2} \Big) \,
		\Gamma\Big( \frac{4-i\omega+i\omega'}{2} \Big) \, \Gamma\Big( \frac{2-i\omega-i\omega'}{2} \Big) \nonumber\\
		& \quad + \, \Gamma\Big( \frac{2+i\omega+i\omega'}{2} \Big) \, \Gamma\Big( \frac{4+i\omega-i\omega'}{2} \Big) \,
		\Gamma\Big( \frac{-i\omega+i\omega'}{2} \Big) \, \Gamma\Big( \frac{2-i\omega-i\omega'}{2} \Big) \nonumber\\
		& \quad + \, \Gamma\Big( \frac{4+i\omega+i\omega'}{2} \Big) \, \Gamma\Big( \frac{2+i\omega-i\omega'}{2} \Big) \,
		\Gamma\Big( \frac{2-i\omega+i\omega'}{2} \Big) \, \Gamma\Big( \frac{-i\omega-i\omega'}{2} \Big) \bigg] \nonumber\\
		=& \ - \, \frac{\pi^2 (\omega^2 - \omega'^2)}{4[\cosh(\pi \omega) - \cosh(\pi \omega')]} \, .
	\end{align}
Hence, the Hamiltonian is
	\begin{align}
		H \, =& \ \int d\omega dk \, \frac{\omega}{2} \bigg[ b_i^R(\omega, k) \, b_i^{R\dagger}(\omega, k) \, + \, b_i^{R\dagger}(\omega, k) \, b_i^R(\omega, k) \ \bigg] \, .
	\label{rindler-hamiltonian}
	\end{align}
Note that the contributions from $F_2$ and $F_3$ exactly cancel in $b^Rb^{R\dagger}$, $b^{R\dagger}b^R$ terms.



\begin{thebibliography}{99}


\bibitem{Mathur:2012np} 
S.~D.~Mathur, \emph{The information paradox: conflicts and resolutions}, \emph{Pramana} {\bf 79} (2012) 1059~[arXiv:1201.2079~[hep-th]].


\bibitem{Braunstein:2009} 
S.~L.~Braunstein, S.~Pirandola, and K.~${\rm \dot{Z}}$yczkowski, \emph{Better Late than Never: Information Retrieval from Black Holes}, \emph{Phys.\ Rev.\ Lett.}\ {\bf 110}, (2013) 101301~[arXiv:0907.1190~[quant-ph]].


\bibitem{Almheiri:2012rt} 
A.~Almheiri, D.~Marolf, J.~Polchinski and J.~Sully, \emph{Black Holes: Complementarity or Firewalls?}, \emph{JHEP} {\bf 1302} (2013) 062~[arXiv:1207.3123~[hep-th]].


\bibitem{Marolf:2013dba} 
D.~Marolf and J.~Polchinski, \emph{Gauge/Gravity Duality and the Black Hole Interior}, \emph{Phys.\ Rev.\ Lett.}\ {\bf 111} (2013) 171301~[arXiv:1307.4706~[hep-th]].


\bibitem{VanRaamsdonk:2010pw} 
M.~Van Raamsdonk, \emph{Building up spacetime with quantum entanglement},
\emph{Gen.\ Rel.\ Grav.}\ {\bf 42} (2010) 2323~[\emph{Int.\ J.\ Mod.\ Phys.\ D} {\bf 19} (2010) 2429]~[arXiv:1005.3035~[hep-th]].


\bibitem{Czech:2012}
B.~Czech, J.~L.~Karczmarek, F.~Nogueira, and M.~V.~Raamsdonk, \emph{Rindler Quantum Gravity}, 
\emph{Class. Quant. Grav.} {\bf 29} (2012) 235025~[arXiv:1206.1323~[hep-th]].


\bibitem{Parikh:2012}
M.~Parikh, and P.~Samantray, \emph{Rindler-AdS/CFT}, [arXiv:1211.7370~[hep-th]].


\bibitem{Emparan:1999}
R.~Emparan, \emph{AdS/CFT Duals of Topological Black Holes and the Entropy of Zero-Energy States}, \emph{JHEP} {\bf 9906} (1999) 036~[arXiv:9906040 [hep-th]].


\bibitem{Aros:2002}
R~.Aros, C.~Martinez, R.~Troncoso, and J.~Zanelli, \emph{Quasinormal modes for massless topological black holes},
\emph{Phys. Rev.} D {\bf 67}, 044014 (2003)~[arXiv:0211024 [hep-th]].


\bibitem{Maldacena:2001kr}
J.~M.~Maldacena, \emph{Eternal black holes in anti-de Sitter}, \emph{JHEP} {\bf 0304} (2003) 021~[hep-th/0106112].


\bibitem{Takahashi:1996zn} 
Y.~Takahasi and H.~Umezawa, \emph{Collective Phenomena} {\bf 2} (1975) 55;\\
Y.~Takahashi and H.~Umezawa, \emph{Thermo field dynamics}, \emph{Int.\ J.\ Mod.\ Phys.}\ B {\bf 10} (1996) 1755.


\bibitem{Israel:1976ur}
W.~Israel, \emph{Thermo field dynamics of black holes}, \emph{Phys.\ Lett.}\ A {\bf 57} (1976) 107.


\bibitem{Schwinger:1960qe} 
J.~S.~Schwinger, \emph{Brownian motion of a quantum oscillator}, \emph{J.\ Math.\ Phys.}\  {\bf 2} (1961) 407.


\bibitem{Keldysh:1964ud} 
L.~V.~Keldysh, \emph{Diagram technique for nonequilibrium processes}, \emph{Zh.\ Eksp.\ Teor.\ Fiz.}\  {\bf 47} (1964) 1515~[\emph{Sov.\ Phys.\ JETP} {\bf 20} (1965) 1018].


\bibitem{Semenoff:1982ev} 
G.~W.~Semenoff and H.~Umezawa, \emph{Functional Methods in Thermo Field Dynamics: A Real Time Perturbation Theory for Quantum Statistical Mechanics},
\emph{Nucl.\ Phys.}\ B {\bf 220} (1983) 196.


\bibitem{Niemi:1983nf} 
A.~J.~Niemi and G.~W.~Semenoff, \emph{Finite Temperature Quantum Field Theory in Minkowski Space}, \emph{Annals Phys.}\ {\bf 152} (1984) 105.


\bibitem{Ojima:1981ma} 
I.~Ojima, \emph{Gauge Fields at Finite Temperatures: Thermo Field Dynamics, KMS Condition and their Extension to Gauge Theories}, \emph{Annals Phys.}\  {\bf 137} (1981) 1.


\bibitem{Maldacena:2013xja} 
J.~Maldacena and L.~Susskind, \emph{Cool horizons for entangled black holes}, \emph{Fortsch.\ Phys.}\  {\bf 61} (2013) 781~[arXiv:1306.0533~[hep-th]].


\bibitem{Avery:2013bea}
S.~G.~Avery and B.~D.~Chowdhury, \emph{No Holography for Eternal AdS Black Holes} [arXiv:1312.3346~[hep-th]].


\bibitem{Mathur:2014dia} 
S.~D.~Mathur, \emph{What is the dual of two entangled CFTs?}~[arXiv:1402.6378~[hep-th]].


\bibitem{Nair:2015} 
V.~P.~Nair, \emph{Thermofield dynamics and Gravity}~[arXiv:1508.00171~[hep-th]].


\bibitem{Hamilton:2006}
A.~Hamilton, D.~Kabat, G.~Lifschytz, and D.~A.~Lowe, \emph{Holographic representation of local bulk operators}, arXiv:0606141 [hep-th].\\


\bibitem{Bousso:2012mh} 
R.~Bousso, B.~Freivogel, S.~Leichenauer, V.~Rosenhaus and C.~Zukowski, \emph{Null Geodesics, Local CFT Operators and AdS/CFT for Subregions},
\emph{Phys.\ Rev.}\ D {\bf 88} (2013) 064057~[arXiv:1209.4641 [hep-th]].


\bibitem{Fradkin:1986ka} 
E.~S.~Fradkin and M.~A.~Vasiliev, \emph{Candidate to the Role of Higher Spin Symmetry}, \emph{Annals Phys.}\  {\bf 177} (1987) 63.


\bibitem{Vasiliev:1990en} 
M.~A.~Vasiliev, \emph{Consistent equation for interacting gauge fields of all spins in (3+1)-dimensions}, \emph{Phys.\ Lett.}\ B {\bf 243} (1990) 378.


\bibitem{Vasiliev:1995dn} 
M.~A.~Vasiliev, \emph{Higher spin gauge theories in four-dimensions, three-dimensions, and two-dimensions},
\emph{Int.\ J.\ Mod.\ Phys.}\ D {\bf 5} (1996) 763~[arXiv:hep-th/9611024].


\bibitem{Bekaert:2005vh} 
X.~Bekaert, S.~Cnockaert, C.~Iazeolla and M.~A.~Vasiliev, \emph{Nonlinear higher spin theories in various dimensions},~[arXiv:hep-th/0503128].


\bibitem{Klebanov:2002ja} 
I.~R.~Klebanov and A.~M.~Polyakov, \emph{AdS dual of the critical O(N) vector model}, \emph{Phys.\ Lett.}\ B {\bf 550} (2002) 213~[arXiv:hep-th/0210114].


\bibitem{Sezgin:2002rt} 
E.~Sezgin and P.~Sundell, \emph{Massless higher spins and holography}, \emph{Nucl.\ Phys.}\ B {\bf 644} (2002) 303~[Erratum-ibid.\ B {\bf 660} (2003) 403]~[hep-th/0205131].


\bibitem{Giombi:2009wh} 
S.~Giombi and X.~Yin, \emph{Higher Spin Gauge Theory and Holography: The Three-Point Functions}, \emph{JHEP} {\bf 1009} (2010) 115~[arXiv:0912.3462~[hep-th]].


\bibitem{Koch:2010cy} 
R.~d.~M.~Koch, A.~Jevicki, K.~Jin and J.~P.~Rodrigues, \emph{$AdS_4/CFT_3$ Construction from Collective Fields},
\emph{Phys.\ Rev.}\ D {\bf 83} (2011) 025006~[arXiv:1008.0633~[hep-th]].


\bibitem{Maldacena:2011jn}
J.~Maldacena and A.~Zhiboedov, \emph{Constraining Conformal Field Theories with A Higher Spin Symmetry},
\emph{J.\ Phys.}\ A {\bf 46} (2013) 214011~[arXiv:1112.1016~[hep-th]].


\bibitem{Kraus:2011ds} 
P.~Kraus and E.~Perlmutter, \emph{Partition functions of higher spin black holes and their CFT duals}, \emph{JHEP} {\bf 1111} (2011) 061~[arXiv:1108.2567~[hep-th]].


\bibitem{Gaberdiel:2012yb} 
M.~R.~Gaberdiel, T.~Hartman and K.~Jin, \emph{Higher Spin Black Holes from CFT}, \emph{JHEP} {\bf 1204} (2012) 103~[arXiv:1203.0015~[hep-th]].


\bibitem{Ammon:2012wc} 
M.~Ammon, M.~Gutperle, P.~Kraus and E.~Perlmutter, \emph{Black holes in three dimensional higher spin gravity: A review}
\emph{J.\ Phys.}\ A {\bf 46} (2013) 214001~[arXiv:1208.5182 [hep-th]].

 
\bibitem{Gaberdiel:2013jca} 
M.~R.~Gaberdiel, K.~Jin and E.~Perlmutter, \emph{Probing higher spin black holes from CFT}, \emph{JHEP} {\bf 1310} (2013) 045~[arXiv:1307.2221~[hep-th]].


\bibitem{Das:2003vw} 
S.~R.~Das and A.~Jevicki, \emph{Large N collective fields and holography}, \emph{Phys.\ Rev.\ D} {\bf 68} (2003) 044011~[hep-th/0304093].


\bibitem{Koch:2014aqa} 
R.~d.~M.~Koch, A.~Jevicki, J.~P.~Rodrigues and J.~Yoon, \emph{Canonical Formulation of $O(N)$ Vector/Higher Spin Correspondence}, [arXiv:1408.4800~[hep-th]].


\bibitem{Mintun:2014gua} 
E.~Mintun and J.~Polchinski, \emph{Higher Spin Holography, RG, and the Light Cone}~[arXiv:1411.3151~[hep-th]]. 


\bibitem{Leigh:2014qca} 
R.~G.~Leigh, O.~Parrikar and A.~B.~Weiss, \emph{Exact renormalization group and higher-spin holography},
\emph{Phys.\ Rev.}\ D {\bf 91} (2015) 026002~[arXiv:1407.4574~[hep-th]].


\bibitem{Jin:2015aba} 
K.~Jin, R.~G.~Leigh and O.~Parrikar, \emph{Higher Spin Fronsdal Equations from the Exact Renormalization Group},~[arXiv:1503.06864~[hep-th]].


\bibitem{Koch:2014mxa}
R.~de Mello Koch, A.~Jevicki, J.~P.~Rodrigues and J.~Yoon, \emph{Holography as a Gauge Phenomenon in Higher Spin Duality},
\emph{JHEP} {\bf 1501} (2015) 055~[arXiv:1408.1255 [hep-th]].
 

\bibitem{Jevicki:2015sla}
A.~Jevicki, and J.~Yoon, \emph{Bulk from Bi-locals in Thermo Field CFT}~[arXiv:1503.08484~[hep-th]].


\bibitem{Belin:2014}
A.~Belin, and A.~Maloney, \emph{A New Instability of the Topological black hole}~[arXiv:1412.0280~[hep-th]].


\bibitem{Rey:2014dpa} 
S.~J.~Rey and V.~Rosenhaus, \emph{Scanning Tunneling Macroscopy, Black Holes, and AdS/CFT Bulk Locality},
\emph{JHEP} {\bf 1407} (2014) 050~[arXiv:1403.3943~[hep-th]].


\bibitem{Papadodimas:2012aq} 
K.~Papadodimas and S.~Raju, \emph{An Infalling Observer in AdS/CFT}, \emph{JHEP} {\bf 1310} (2013) 212~[arXiv:1211.6767~[hep-th]].


\bibitem{Papadodimas:2013jku} 
K.~Papadodimas and S.~Raju, \emph{State-Dependent Bulk-Boundary Maps and Black Hole Complementarity},
\emph{Phys.\ Rev.\ D} {\bf 89} (2014) 086010~[arXiv:1310.6335~[hep-th]].


\bibitem{Kraus:2002iv} 
P.~Kraus, H.~Ooguri and S.~Shenker, \emph{Inside the horizon with AdS / CFT}, \emph{Phys.\ Rev.}\ D {\bf 67} (2003) 124022~[hep-th/0212277].


\bibitem{Hawking}
S.~W.~Hawking, \emph{Particle Creation by Black Holes}, Commun.\ Math.\ Phys. {\bf 43} 199-220 (1975).


\bibitem{Unruh}
W.~G.~Unruh, \emph{Notes On Black Hole Evaporation}, Phys.\ Rev.\ D {\bf 14}, 870 (1976).


\bibitem{Didenko:2009td} 
V.~E.~Didenko and M.~A.~Vasiliev, \emph{Static BPS black hole in 4d higher-spin gauge theory},
Phys.\ Lett.\ B {\bf 682}, 305 (2009)\ [Phys.\ Lett.\ B {\bf 722}, 389 (2013)]\ [arXiv:0906.3898 [hep-th]].


\bibitem{Longhi}
P.~Longhi, and R.~Soldati, \emph{The Unruh effect revisited}, \emph{Phys. Rev.} D {\bf83} 107701, (2011)~[arXiv:1101.5976 [hep-th]].


\bibitem{Jeffrey}
A.~Jeffrey, and D.~Zwillinger, \emph{Table of Integrals, Series, and Products}, Academic Press; 7 edition (March 9, 2007).












\end{thebibliography}
\end{document}